\documentclass[aps,prb,a4paper,twocolumn,floatfix,showpacs,preprintnumbers]{revtex4-1}

\usepackage{mathrsfs}
\usepackage{amsmath}
\usepackage{amssymb}
\usepackage{graphicx}
\usepackage{color}

\begin{document}

\title{Andreev reflection tuned by magnetic barriers in superconductor contacted graphene}
\author{Hengyi Xu}\affiliation{Condensed Matter Physics Laboratory, Heinrich-Heine-Universit\"at,
Universit\"atsstr.1, 40225 D\"usseldorf, Germany}
\author{T. Heinzel}\affiliation{Condensed Matter Physics Laboratory, Heinrich-Heine-Universit\"at,
Universit\"atsstr.1, 40225 D\"usseldorf, Germany}

\date{\today }

\begin{abstract}
The Andreev reflection of the normal state-superconductor junction both in monolayer and bilayer graphene with a single magnetic barrier is investigated by means of the Green's function formalism. Within the tight-binding model, we study the direction-dependent Andreev reflection of two-dimensional graphene-superconductor junctions in the specular and retro-reflection regimes. The presence of a magnetic barrier close to the superconducting hybrid junction introduces a rich phenomenology. Such a barrier is capable of tuning the preferred angles of incidence for the Andreev reflection. In particular, it can enhance the specular reflection probability for certain angles of incidence in bilayer-based hybrid structures. When transmission is permitted, the Andreev reflection manifests itself in isolated peaks and asymmetric resonances associated with offsets and Fano-type oscillations in the transmission, respectively. Moreover, Fabry-P\'{e}rot oscillations in the Andreev reflection due to the interior scattering inside the magnetic barrier may appear. The impacts of magnetic barriers on the monolayer and bilayer hybrid interfaces are furthermore studied by calculating the differential conductances within the Blonder-Tinkham-Klapwijk formula.

\end{abstract}

\maketitle

\section{Introduction}
A dissipative electric current is converted into a dissipationless supercurrent by the celebrated Andreev retro-reflection at the normal-superconducting (N-SC) interface where an electron is reflected into a hole which retraces the path of the incident electron. \cite{BeenakkerRMP1997} In graphene, it has been suggested that due to its peculiar band structure, the Andreev reflection (AR) can be specular for excitation energies above the Fermi energy. The specular AR in graphene differs from the usual retro-reflection with clear experimental signatures. \cite{BeenakkerPRL2006} Contrary to two-dimensional graphene, the AR in graphene nanoribbons (GNR) is sensitive to the ribbon width and the pseudoparity of quantum states. \cite{RainisPRB2007} Besides the monolayer graphene (MLG)-superconductor junction, the bilayer graphene (BLG)-superconductor junction has been studied by Ludwig, and a remarkable suppression of the differential conductance at voltages just below the superconducting gap was found. \cite{LudwigPRB2007}

Beyond a simple N-SC junction, a number of extended hybrid structures have also been proposed and studied, like for example the graphene-insulator-SC \cite{Bhattacharjee2006} and  the SC-graphene-SC systems \cite{CuevasPRB2006} as well as a multi-terminal system of graphene-SC junctions where the subgap structure and crossed Andreev reflection were discussed \cite{CayssolPRB2008,HaugenPRB2010}. In particular, the transport properties through the graphene-SC junctions in the presence of magnetic fields have been studied. The distinct transport regimes are recognizable in the phase diagram. \cite{PradaPRB2007}

On the other hand, the intriguing properties of graphene with normal contacts have been extensively investigated both experimentally and theoretically. \cite{CastroNeto2009,Novoselov2005,Han2007,Katsnelson2006,Tworzydlo2006,Pereira2006} Graphene exhibits a variety of exotic transport behaviors, such as the anomalous integer and fractional quantum Hall effect and minimal conductivity \cite{Novoselov2005,Zhang2005}, as well as, an extremely high mobility due to the fact that the Dirac fermions can pass perfectly through classically forbidden region (Klein tunneling) \cite{Morozov2008,Sarma2011,Peres2011,Xu2011PRB}. This effect, however, impedes the the confinement of Dirac fermions for the purpose of future applications. It has been shown that a local inhomogeneous magnetic field,  i.e. a magnetic barrier (MB), is able to restrict massless Dirac electrons effectively. \cite{MartinoPRL2007,Anna2009} Moreover, the electronic tunneling through various magnetic barrier geometries in graphene has been studied systematically by Masir et al. \cite{MasirPRB2008,Masir2008APL} These studies have paved the way for constructing direction-dependent wave vector filters based on the MB which is viewed as a key element for graphene-based electronics. \cite{Zhai2010} In view of the fact that superconducting contacts as well as MBs in graphene give rise to numerous novel transport phenomena, it appears to be a natural next step to combine these two components and explore the resulting transport properties. In terms of fundamental physics, such a study will deepen our insight into the impact of MBs on the interplay between superconductivity and relativistic dynamics. Furthermore, it will also provide a guideline for possible future experimental steps in this direction.

In this paper, we investigate the AR both at the MLG-SC and the BLG-SC junctions with a single MB by utilizing the real-space Green's function formalism. The angular dependencies of the AR at the electron energy lower or slightly higher than the superconducting energy gap are calculated. The response of the hybrid junction for a single MB is examined in different transport regimes. Also, the differential conductance of the NSC interface with a MB closeby is considered to facilitate comparison to experiments.

This paper is organized as follows. In Sec. II we sketch the considered geometry and introduce the formalism. This is followed by the presentation and discussion of the numerical results in Sec. III. The summary and conclusions constitute Sec. IV.

\section{Model and Formalism}
\begin{figure}[tbp]
\includegraphics[keepaspectratio,width=\columnwidth]{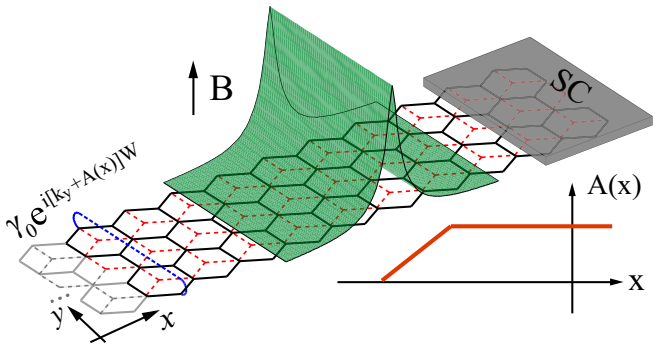}
\caption{(Color online) Schematic geometry of the hybrid structure under consideration in monolayer or bilayer graphene. The central graphene device, where a local inhomogeneous magnetic field
is applied, is connected by two semi-infinite graphene leads. The right lead is contacted by a superconducting metal to induce superconductivity. The tight-binding model satisfies boundary conditions to model two-dimensional graphene. Lower panel: A typical vector potential profile for a single magnetic barrier.}
\label{fig0}
\end{figure}

We consider the two-terminal hybrid graphene-SC geometry sketched in Fig. \ref{fig0}. The left and right leads are made of the semi-infinite metallic and superconducting graphene, respectively. In the central region, a single magnetic barrier is applied to the normal-conducting graphene. This system can be described by the Dirac-Bogoliubov-de Gennes (DBdG) Hamiltonian with finite and vanishing pair potentials $\Delta=\Delta_0e^{i\varphi}$ corresponding to the superconducting and normal-states graphene, respectively. In this work we consider the $s$-wave symmetric pair potential, i.e., $\Delta_0$ is real and the superconducting phase $\varphi=0$.

The easiest way to induce superconductivity in graphene relies on the proximity effect by superimposing a superconductor (SC) on graphene as shown in the upper panel of Fig. \ref{fig0}. Recent experiments have observed unambiguously superconductivity in graphene via this mechanism. \cite{Heersche2007,Heersche200772,Allain2012} The details of the induction of superconductivity and the lattice matching between graphene and the SC are of no interest for our purpose.

We make use of the tight-binding (TB) approach for the description of the transport calculations.
The DBdG Hamiltonian for MLG in TB model reads
\begin{eqnarray}
\mathcal {H}_{\mathrm{MLG}}=&&\sum_i \left[(V_i-E_F)\sigma^z +\frac{\Delta}{2}\sigma_+ +\frac{\Delta^*}{2}\sigma_-\right]c^\dag_i c_i  \nonumber\\
&-& \gamma_0\sum_{\langle i,j\rangle}\sigma^zc^\dag_i c_{j}\label{hmlg}
\end{eqnarray}
where $c_i^\dag(c_i)$ is the creation (annihilation) operator on the site $i$. The first sum runs over all the atomic sites with on-site electrostatic potential $V_i$. $E_F$ is the Fermi energy, and the $2\times 2$ Pauli matrices $\vec{\sigma}=(\sigma^x,\sigma^y,\sigma^z)$ act on the electron-hole space. $\sigma_{\pm}=\sigma^x\pm i\sigma^y$ are the raising and lowering operators, respectively. The pair potential $\Delta(x)$ describes the coupling between electrons and holes in superconducting states, which vanishes in the graphene normal state. The second sum runs over all the pairs of neighboring atomic site $\langle i,j\rangle$ with the hopping energy $\gamma_0\approx 3$ eV. All states are spin-degenerate.

The Hamiltonian describing the BLG in Bernal stacking is written as \cite{Xu2009PRB,Xu2010PRB}
\begin{equation}
\mathcal{H}_{\mathrm{BLG}}=\sum_{\nu=1,2}\mathcal{H}_{\mathrm{MLG}}^{(\nu)}-\gamma_1\sum_i(c_{1,i}^\dag d_{2,i}+\mathrm{H.c.})\label{hblg}
\end{equation}
where $\mathcal{H}_{\mathrm{MLG}}^{\nu}$ is the Hamiltonian of the $\nu$-th graphene layer given by Eq. (\ref{hmlg}), $c_{\nu,i}^\dag(d_{\nu,i})$ is the creation (annihilation) operator at sublattice $A(B)$, in the layer $\nu=1,2$. $\gamma_1=0.39$ eV is the coupling energy between sublattice $B$ and $A$ in different graphene layers, which is the most relevant for the low-energy band of BLG. Other interlayer coupling energies, in particular $\gamma_3$ and $\gamma_4$ leave the basic phenomenology unaltered and are therefore not considered here. \cite{Dresselhaus2002,Mccann2006}

The transport properties of the two-terminal system is calculated by means of the real-space Green's function formalism which has been discussed in detail in Refs. [\onlinecite{WimmerPhD},\onlinecite{Xu2008PRB}]. Here, we merely give the most relevant formalism for the present calculations. The total retarded Green's function is calculated by the recursive algorithm and connected to the two leads by
\begin{equation}
G^R(\epsilon)=(E_F+\epsilon-\mathcal{H}_D-\Sigma_L-\Sigma_R)^{-1}
\end{equation}
where $\epsilon$ is the excitation energy relative to the Fermi energy $E_F$. $\mathcal{H}_D$ is the Hamiltonian of the central device region including scattering, the self-energies $\Sigma_{L(R)}$ accounts for the effects of the semi-infinite left and right leads, respectively. The right lead is superconducting, while the left lead where the wave enters is always in the normal state. The transmission amplitude for a particle $p$ in mode $n$ entering from the left lead and exiting into the right lead as the particle $p'$ in mode $m$ is given by
\begin{equation}
t_{rl,mn}^{p'p}=\frac{i}{\hbar}\frac{1}{\sqrt{|v_n^{p}||v_m^{p'}|}} (\phi_{m}^{(r)})^\dagger\Gamma_{r}^{p'} G^R_{N+1,0}\Gamma_l^p\phi_{n}^{(l)}
\end{equation}
where $G^R_{N+1,0}=\langle N+1| G^R(\epsilon)|0\rangle$ is the Green's function element in the real-space with $0$ and $N+1$ denoting the first slice in the left and right leads, respectively, $\Gamma_{l(R)}$ is the spectral matrix given by
$\Gamma_{L(R)}\equiv i(\Sigma_{l(R)}-\Sigma_{l(R)}^\dagger)$. $\phi_{n(m)}$ is the eigenfunction of the Bloch state $n(m)$ with the group velocity $v_{n(m)}$.

The reflection amplitude of a particle $p$ in mode $n$ in the left lead $l$ to a particle $p'$ which may be either an electron ($e$) or a hole ($h$) of the mode $m$ in the same lead $l$ reads
\begin{eqnarray}
r_{ll,mn}^{p'p}=\frac{1}{\hbar}\frac{1}{\sqrt{|v_n^{p}||v_m^{p'}|}}&&\left(i (\phi_{m}^{(l)})^\dagger\Gamma_{l}^{p'} G^R_{0,0}\Gamma_l^p\phi_{n}^{(l)}\right.\nonumber\\
&&\left.-(\phi_{m}^{(l)})^\dagger\Gamma_{l}^{p'}\phi_{n}^{(l)}\delta_{p'p}\right)
\end{eqnarray}
where the Green's function element $G^R_{0,0}=\langle 0| G^R(\epsilon)|0\rangle$. Apparently, $r^{he}$ and $r^{ee}$ denote the Andreev and normal reflection amplitudes, respectively.

The corresponding reflection $R$ and transmission $T$ probabilities are given by
\begin{equation}
R(T)^{p'p}=\sum_{m,n}\left|r(t)^{p'p}_{mn}\right|^2.\label{reflt}
\end{equation}
Henceforth we use $R^A$ and $R^N$ to denote the Andreev and normal reflections, respectively.

By imposing periodic boundary conditions on the transverse ($y$-) direction (thereby wrapping the chain into a loop and removing its ends), the Green's function technique is able to treat the transport properties of 2D graphene layers. In our calculations, this is implemented by connecting the upmost and downmost carbon atoms of the GNR of a width of $W$ by the hopping energy $\gamma_0e^{ik_yW}$ with the wave vector $k_y$ in the y-direction. This technique is rather efficient for dealing with the transport properties of the 2D graphene. \cite{LiuPRB2012}

To incorporate the perpendicular magnetic field in this model, we start with the Dirac equation instead of Peierls' substitution usually applied within the TB model. In more detail, the momentum in the presence of magnetic fields becomes $\mathbf p+e\mathbf A(x)$. By using the Landau gauge, the vector potential is given by $\mathbf A=(0, B(x)x, 0)$ with $B(x)$ being the inhomogeneous magnetic field along the $x$-direction. In our calculations, the MB is parameterized by its length $L$ and strength $B$, i.e. a rectangular barrier, which generates a vector potential profile shown in the lower panel of Fig. \ref{fig0}. A more realistic barrier shape as implemented in experiments  \cite{Tarasov2010} is shown in the main figure, but the shape modifies the vector potential profile only slightly in its incremental part and does not affect our numerical results significantly. Consequently, the MB is incorporated by rewriting the periodic hopping energy as $\gamma_0e^{i[k_y+A_y(x)]W}$ with $A_y(x)=\int_{-\infty}^xB(x)dx$.

The differential conductance of the system is calculated according to the Blonder-Tinkham-Klapwijk formula \cite{Blonder1982} as
\begin{equation}
g=\frac{\partial I}{\partial V}=g_0\int_{-\pi/2}^{\pi/2}(1-R^N+R^A)\cos\alpha d\alpha,
\end{equation}
where $g_0$ is the ballistic differential conductance for either MLG or BLG and includes the spin and valley degeneracies.
The normal reflection $R^N$ and Andreev reflection $R^A$ are calculated from Eq. (\ref{reflt}).

\section{Results and Discussion}
In this section we present and discuss the numerical results of our transport calculations. The hybrid structures in both MLG and BLG are investigated. We consider two different regimes, i.e. the subgap regime with incoming energy $\epsilon<\Delta_0$ and the open regime with $\epsilon>\Delta_0$.

\subsection{subgap regime $\epsilon<\Delta_0$}

\begin{figure}[tbp]
\includegraphics[scale=0.8]{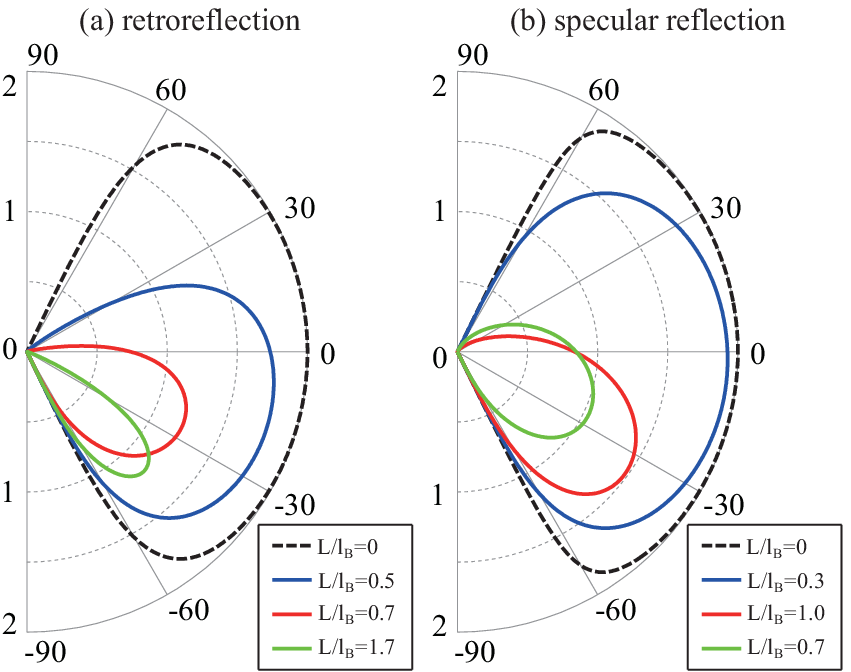}
\caption{(Color online) The angular dependence of the Andreev reflection for monolayer graphene-SC hybrid junction with various MBs in the subgap regime. (a) The retro-reflection with $\epsilon/E_F=0.05$ and $E_F/\Delta_0=10$, $\alpha_c\approx 65^\circ$. The blue, red, and green curves from exterior to interior respectively correspond to $kl_B=1.3$, $0.9$, and $1.3$. (b)  The specular reflection with $\epsilon/E_F=19$ and $E_F/\Delta_0=0.05$, $\alpha_c\approx 64^\circ$. The blue, red, and green curves from exterior to interior respectively correspond to $kl_B=1.7$, $1.2$, and $0.9$. For MLG, $\hbar v_Fk=E_F+\epsilon$ with $v_F$ the Fermi velocity.}
\label{fig1}
\end{figure}

We first focus on the subgap regime. The AR in this regime can be further identified as the retro- ($\epsilon<E_F$) and specular ($\epsilon>E_F$) reflection depending upon the excitation energy relative to the Fermi energy. Fig. \ref{fig1} shows the AR as a function of the angles of incidence with increasing of the strength of MBs. The AR vanishes if the angle of incidence is beyond a critical angle $\alpha_c$ which is given by
\begin{equation}
\alpha_c=\arcsin\left(\frac{|\epsilon-E_F|}{\epsilon+E_F}\right)^{1/\nu},\label{alphac}
\end{equation}
with $\nu=1,2$ for monolayer and bilayer graphene, respectively. \cite{BeenakkerPRL2006,LudwigPRB2007} Inspection from the dispersion relation, this is because a gap is opened in the hole band for small longitudinal momentums. The numerical results coincide well with the Eq. (\ref{alphac}). In the absence of MBs, a common feature for the retro-reflection and specular reflection in MLG is the unity reflection at normal incidence ($\alpha=0$) due to the conservation of chirality in contrast to the usual NSC junctions. Namely, the reflection from an electron to a hole at the graphene-SC interface does not involve the conversion of sublattices, whereas the normal reflection $R^N$ does involve scattering between sublattices at $\alpha=0$. \cite{BeenakkerPRL2006} Note that the AR probability of $2$ in our case is due to the valley degeneracy.

The magnetic field filters the wave vectors prior to the electron approaching the the NSC interface, making the AR probabilities asymmetric with respect to the normal incidence owing to the time-reversal symmetry breaking. As the strength of the MB increases, the suppression of AR by MBs becomes more prominent. For MLG hybrid structures, the retro- and specular reflections behave similarly.

\begin{figure}[tbp]
\includegraphics[scale=0.8]{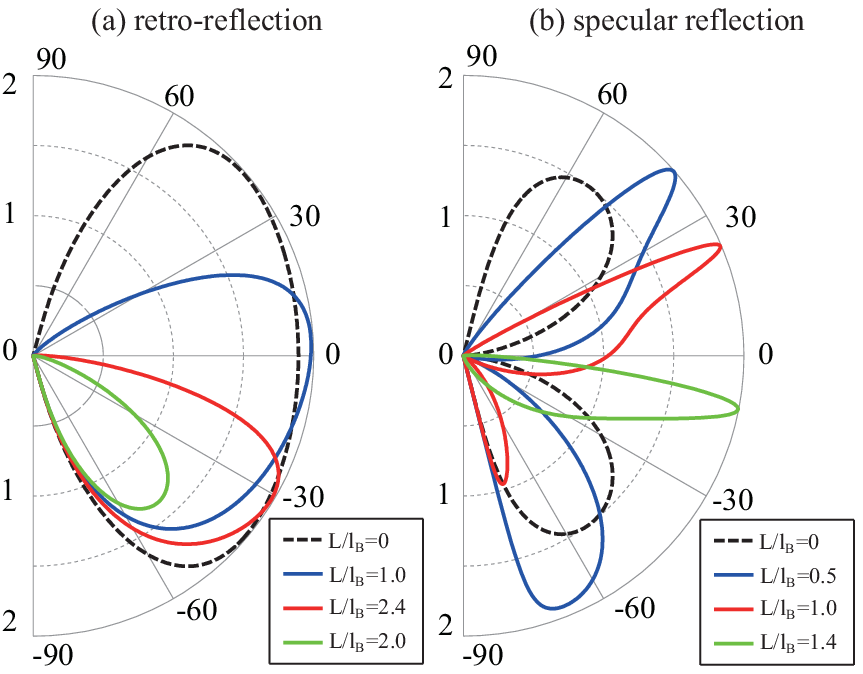}
\caption{(Color online) The angular dependence of the Andreev reflection for bilayer graphene-SC hybrid junction at various MBs in the subgap regime. (a) The retro-reflection with $\epsilon/E_F=0.03$ and $E_F/\Delta_0=2$, $\alpha_c\approx 70^\circ$. The blue, read, and green curves from exterior to interior respectively correspond to $kl_B=2$, $2$, and $1.5$. (b) The specular reflection with $\epsilon/E_F=37$ and $E_F/\Delta_0=0.02$, $\alpha_c\approx 71^\circ$. The blue, red, and green curves from exterior to interior respectively correspond to $kl_B=1.8$, $1.8$, and $1.3$. For BLG, $\hbar^2 v_F^2k^2/\gamma_1=E_F+\epsilon$ with $v_F$ the Fermi velocity.}
\label{fig2}
\end{figure}

In comparison to the case of MLG, the retro-reflection and specular reflection in the BLG system exhibit much more distinct features as shown in Fig. \ref{fig2}. Considering that the dispersion of BLG is not perfectly parabolic, the numerical values of $\alpha_c$ deviate slightly from those from Eq. (\ref{alphac}). More significantly, the retro-reflections at $\epsilon<E_F$ always take place with a certain probability at normal incidence, while the specular reflections at $\epsilon>E_F$ always show zero probabilities corresponding to perfect normal reflections, i.e. $R^N=1$. This is consistent with the theoretical results based on Dirac theory and can be interpreted by the compatibility of the spinor wave functions of excitations in graphene. \cite{LudwigPRB2007} To be more specific, the wave function of an incident electron in the sublattice basis is orthogonal to that of the reflected hole in the valence band so that specular AR is forbidden at $\alpha=0$. This feature persists in the presence of MBs, while the angular dependence of the AR is bent downwards.

Furthermore, the MBs may raise the AR probability up to unity for certain directions, particularly for specular reflections. The two wings of the specular AR are bent and expanded for the smallest magnetic field. As the magnetic field increases, the lower wing shrinks, while the upper wing persists. The enhancement in the upper wing of the AR can be understood as follows. The electrons may undergo successive scattering inside a resonant cavity formed by the MB and the N-SC interface. Each of the cycles can build up the probability of conversion from electrons to holes. This mechanism and the filtering effect of MBs coexist and compete each other in all cases. However, in other three cases (retro- and specular AR in MLG, and retro-AR in BLG), the AR probabilities nearly saturate for most of incoming angles such that the increments in ARs are negligible. In the time-reversed N-SC junctions, elastic scattering at the hybrid interface and the random impurities produces a constructive interference between the incident electron and the retro-reflected hole resulting in an enhancement in the differential conductance. \cite{Wees1992,Marmorks1993,Giazotto2001} This is known as the ``reflectionless tunneling'' and possibly responsible for the zero-bias conductance peak in graphene hybrid junction. \cite{Popinciuc2012}  However, it is not a major effect here since the specular reflection and the magnetic field spoil the phase conjugation.

\subsection{Open regime $\epsilon>\Delta_0$}

\begin{figure}[tbp]
\includegraphics[scale=0.55]{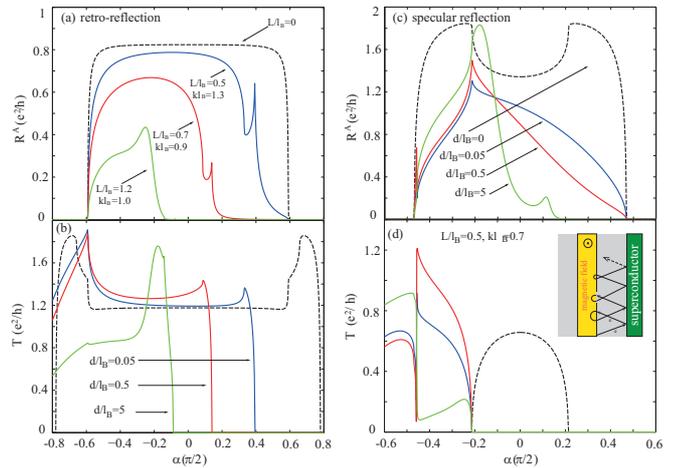}
\caption{(Color online) The Andreev reflection and corresponding transmission probabilities as a function of angles of incidence $\alpha$ in MLG with various MBs at $\epsilon>\Delta_0$. (a) The retro-reflection regime with  $\epsilon/E_F=0.11$ and $E_F/\Delta_0=10$ with the corresponding transmission (b). (c) The specular reflection regime with $\epsilon/E_F=5.1$, and $E_F/\Delta_0=0.2$ for various $d$ with the corresponding transmission (d). The parameters for the MB in this case are fixed as $L/l_B=0.5$ and $kl_B=0.7$. }
\label{fig3}
\end{figure}
We now consider the case of incident electrons with an energy $\epsilon$ slightly above the superconducting gap. In this case, normal transmission emerges and the AR processes are severely suppressed. In Fig. \ref{fig3} we show the Andreev reflection with different MBs and $d$ as a function of angles of incidence $\alpha$. The ARs without MBs are also plotted (the dashed curve) for comparison. Stronger MBs confine the AR successively towards a more negative and narrower regime of angles of incidence because of the filtering effects of the MBs. In Fig. \ref{fig3} (a), the distance $d$ between the MB and the SC interface is also varied. An AR peak appears on the right hand side of the curve and diminishes as $d$ increases. The local AR minimum nearby the peak corresponds to the transmission ``peak" through the N-SC junction as shown in Fig. \ref{fig3}(b), which is a remnant of the protrusion in the transmission without MBs (dashed curve).

Fig. \ref{fig3}(c) shows the AR probability in the specular regime with $\epsilon>E_F$, where we focus on the influence of $d$ on the AR with a constant MB. Compared with the retro-reflection, a concave specular AR shows up around $\alpha=0$, accompanied by a convex-shaped transmission as indicated by the dashed curve in Fig. \ref{fig3}(d).  In the presence of MBs, the overall AR is suppressed except a small angular interval for the largest $d$. Interestingly, there is a clear crossover around $\alpha=-0.1$ (in units of $\pi/2$) separating the specular AR into two regions. In the right region ($\alpha\gtrsim-0.1$), the  transmissions are completely suppressed for all MBs. A possible scenario here is that the electrons are multiply scattered between the MB and the N-SC interface (see the inset of Fig. \ref{fig3}(d) for an example), which compensates the loss in the AR due to the prior scattering of the MB to some extent. Therefore, the AR probability drops as the separation $d$ between the MB and the superconducting lead increases. For the angles of incidence $\alpha\lesssim-0.1$, it crosses over to another region, where the transmission emerges and Fano-type resonances can be observed as shown in Fig. \ref{fig3}(d). What happens here is that, the electron acquires an additional phase after the multiply scattering between the MB and the SC, then transmits into the superconducting lead. The additional phase results in aymmetric line shapes in terms of interference phenomena. \cite{Miroshnichenko2010} Correspondingly, the resulting reflection resonances can be recognized at the angle $\alpha=-0.47$.

\begin{figure}[tbp]
\includegraphics[scale=0.55]{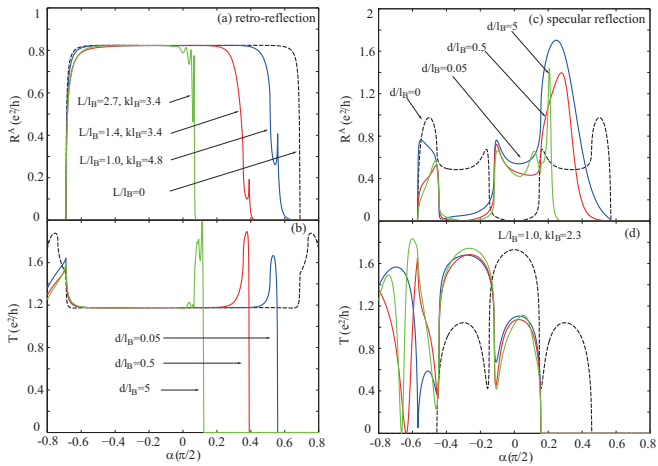}
\caption{(Color online) The Andreev reflection probabilities as a function of angles of incidence $\alpha$ in BLG with various single MBs at $\epsilon>\Delta_0$. (a)  The retro-reflection regime with  $\epsilon/E_F=0.11$ and $E_F/\Delta_0=10$ with the corresponding transmission (b). (b) The specular reflection regime with $\epsilon/E_F=4.2$, and $E_F/\Delta_0=0.25$ for various $d$ with the corresponding transmission (d). The parameters for the MB in this case are fixed as $L/l_B=1.0$ and $kl_B=2.3$.}
\label{fig4}
\end{figure}

In Fig. \ref{fig4}(a), it is shown that the behavior of retro-AR of BLG-SC junctions is very similar to that of the MLG-SC junction. For $d=5l_B$, the reflected holes may undergo interior multiple scattering inside the MB, thereby leading to the Fabry-P\'{e}rot-type oscillations which also appear in the transmission of GNRs subject to a single MB and carbon nanotubes with normal contacts. \cite{Xu2008PRB,Liang2001} For the specular reflection, the MB shifts the reflection minima to negative angles of incdience and raises the reflection probability with the angles of incidence $\alpha\gtrsim 0.2$ markedly as in the subgap regime. The growth of AR gets less prominent, arising from suppression of successive reflections both from the MB and the SC interface as $d$ increases. For $\alpha\lesssim 0.2$, a similar crossover to another region with nonzero normal transmissions can be identified.

\subsection{Differential conductance}
\begin{figure}[tbp]
\includegraphics[scale=0.8]{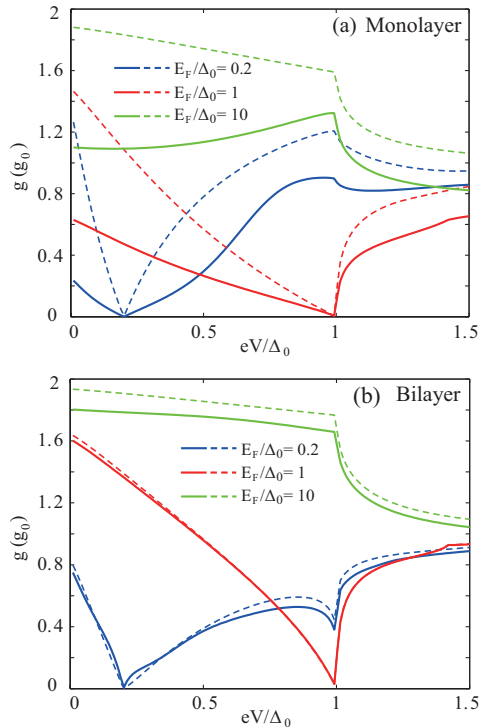}
\caption{(Color online) The differential conductance with a single MB as a function of applied bias for various $E_F/\Delta_0$ ratios. (a) $k_Fl_B=0.24,0.12,1.2$ and $L/l_B=0.5, 0.2, 0.5$ for MLG. Here $\hbar v_Fk_F=E_F$. (b) $k_Fl_B=2,1.4,4.5$ and $L/l_B=0.5, 0.2, 0.5$ for BLG. Here $\hbar^2v_F^2k_F^2/\gamma_1=E_F$. The dashed curves correspond to the cases without MBs. }
\label{fig5}
\end{figure}

We now turn to a discussion of the differential conductance $g$ of N-SC junctions. Fig. \ref{fig5}(a) shows $g$ with various $E_F/\Delta_0$ for MLG hybrid junctions with and without MBs. The conductance drops to zero at $eV=\Delta_0$ since the hole states are unavailable at this voltage. The similar feature can also be found in the case of BLG. The effect of MBs exhibits by reducing the overall differential conductance for all applied voltages except at the singular points, albeit the angular dependence of the AR has been washed out by the angular integration.

The results for the BLG case are shown in Fig. \ref{fig5}(b). By applying the identical MBs on the BLG-SC junction for different $E_F/\Delta_0$, we find that the MBs in BLG hybrid structures are more transparent than those in MLG. Particularly, for $E_F/\Delta_0=0.2$, the conductance with a MB can be slightly higher than that without a MB, because the MB enhanced the Andreev reflections primarily in the specular regime. For $E_F/\Delta_0\ge 1$, the specular regimes are already outside the subgap regime and enter the region where the transmissions are dominant. As a result, the effect is less pronounced. Moreover, a conductance dip, which arises from the zero AR at the normal incidence in the specular regime, shows up at the voltage just below the superconducting energy gap. This conductance dip is also found in the analytical result based on the Dirac equation \cite{LudwigPRB2007}.

\section{Summary and conclusions}
In summary, we have studied the Andreev reflection at normal-conducting-superconduction graphene interfaces with magnetic barriers closeby. The magnetic barrier constrains the permissible angles of incidence, thereby leaving the Andreev reflection asymmetric with respect to the normal incidence, in contrast to electrostatic barriers. In the subgap regime, the magnetic barrier in bilayer graphene shows a quite different behavior for the retro- and specular Andreev reflection compared with the monolayer hybrid junction. In particular, the magnetic barrier can enhance the Andreev reflection for specific angles of incidence due to the multiply reflection of electrons between the magnetic barrier and normal state - superconducting interface.

In the open regime where the transmission is dominant, these hybrid junctions show a rich transport phenomenology. For the retro-reflection, the barrier deflects the propagation of the electron such that the transmission protrusion penetrates into the Andreev reflection regime,  which is manifested by Andreev reflection ``peaks". The multiple scattering of electrons inside the barrier may be significant at small angles of incidence and lead to Fabry-P\'{e}rot interference. In the case of specular reflection, the Andreev reflection can be identified into different transport regions, in which Fano-type resonances may appear.

The magnetic barrier strongly suppresses the differential conductance of monolayer hybrid structures, whereas it shows rather transparency in bilayer. One reason is the nearly parabolic dispersion of bilayer. In graphene contaced by a normal metal, electrons with a parabolic spectrum have shown robustness to magnetic barriers compared to those with the linear spectrum. Also, the magnetic barrier further intensifies the conversion from electrons to holes at the bilayer hybrid interface in the specular regime. In addition, as a result from the Andreev reflection minimal at the normal incidence in the specular regime, the differential conductance of the bilayer junction shows a dip just below the energy gap, which preserves in the presence of magnetic barriers.

The phenomenology of the Andreev retro-reflection can be viewed as an electronic analogue of optical phase conjugation. \cite{kulik00} The presence of the magnetic barrier strongly modifies the preferable direction of the incident electron, and offers a possibility to build a wave-vector-dependent phase-conjugating device.

\acknowledgments
H.X. and T.H. acknowledge financial support from Heinrich-Heine-Universit\"{a}t D\"{u}sseldorf. H.X is greatly thankful to Dr. Ming-Hao Liu for the stimulating discussions and correspondence.

%\bibliography{myref}
%\bibliography{LSHeinzel_Jun27}
%\begin{thebibliography}{99}
%\end{thebibliography}
%

\end{document}